\documentclass[aps,prl,twocolumn,groupedaddress]{revtex4}
\usepackage[dvips]{graphicx}
\begin{document}


\title{Microstructure of neat alcohols}


\author{Aur\'elien Perera$^{1}$ , Franjo Sokoli\'{c}$^{2}$ and Larisa Zorani\'{c}$^{1}$ }
\affiliation{$^{1}$ Laboratoire de Physique Th\'eorique de la Matière Condens\'ee (UMR
CNRS 1600), Universit\'e Pierre et Marie Curie, 4 Place
Jussieu, F75252, Paris cedex 05, France. \\
$^{2}$Laboratoire de Spectrochimie Infrarouge et Raman (UMR
CNRS 8516), Centre d'Etudes et de Recherches Lasers et
Applications, Universit\'edes Sciences et Technologies
de Lille, F59655 Villeneuve d'Ascq Cedex, France.
}


\date{\today}

\begin{abstract}
Formation of microstructure in homogeneous associated liquids is analysed through
the density-density pair correlation functions, both in direct and
reciprocal space, as well as an effective local one-body density function.
This is illustrated through a molecular dynamics study of two neat alcohols, 
namely methanol and \emph{tert}-butanol, which have a rich microstructure: 
chain-like molecular association for the former and micelle-like for the latter. 
The relation to hydrogen bonding interaction is demonstrated.
The apparent failure to find microstructure in water -a stronger hydrogen
bonding liquid- with the same tools, is discussed.
\end{abstract}

\maketitle

Liquids are generally thought as macroscopically homogeneous when
they are considered far from phase transitions and interfacial regions.
From a statistical mechanical point of view, homogeneity is expressed 
by the fact that the order parameter, in this case the one body density,
which formally depends on both the position and orientation of
a single particle $1$ (as in a crystal or a liquid crystal, for example),
is a constant throughout the sample: $\rho^{(1)}(1)=\rho=N/V$, where 
N is the number of particles per volume V. As a consequence, the microscopic
description of the structure of a neat liquid starts from the two-body
density function $\rho^{(2)}(1,2)=\rho^{(1)}(1)\rho^{(1)}(2)g(1,2)$
that expresses the density correlations between particles 1 and 2,
and reduces in this case to $\rho^{2}g(1,2)$ , where $g(1,2)$ is
the pair distribution function. Associated liquids, such as water
and alcohols, for example, belong to a special class because of the
particularity of the hydrogen bonding (HB) 
that is highly directional, and tend to enhance the structure the
liquid locally. One particularly interesting example of this phenomena
is the microheterogeneous nature of aqueous mixtures, which has attracted
a recent upsurge of interest
\cite{soperNAT,bates,guoSpectro,soperPERCO,aupKB,nico,smithMETH}.\\
Perhaps the most remarkable reported fact 
is that water-methanol
mixtures show local immiscibility at microscopic level, while being
miscible at macroscopic level\cite{soperNAT,soperPERCO}.
In order to appreciate this result it is interesting to compare it
to microemulsions where bicontinuous phases are usually observed,
and micro-immiscibility operates
with domain sizes ranging from 100 nanometers to few micrometers,
while those mentioned here are around few nanometers-that is about
few molecular diameters. In addition, it is important to note that
bicontinuous phases in microemulsions arise \emph{after} a phase transition
has occurred from disordered to ordered phase; while in the former case, we are
still in a genuinely homogeneous and disordered liquid phase. From
these facts, microheterogeneity in aqueous mixtures can be considered
as both obvious and mysterious, obvious because 
the mechanism behind it is the strong directionality of the hydrogen
bonding, and mysterious because of the existence of \emph{stable}
micro-immiscibility of water and solute in a macroscopically homogeneous
sample. In contrast to the situation for mixtures, neat water do not
seem to exhibit any micro phase separation between 2 types of water
-tetrahedraly ordered and disordered local regions \cite{frank}
Such a liquid-liquid phase coexistence is recurrent theme in the
study of water \cite{debenedetti,stanleyWaterMyst},
despite the fact that neither experimental results nor computer
simulations have confirmed the validity of this idea. Neat water seems
to behave more like a random network \cite{stanleyPERCO}, with no apparent
particularity in its local distribution, \emph{in average}.\\
In this report, we show by simulations that weak hydrogen bonding liquids,
such as alcohols, have a rich microstructure, that the stronger hydrogen
bonding water do not seem to possess. It is found herein that
neat alcohols are microheterogeneous, since they tend to develop distinct
local microstructures, which depend on the geometry of the constitutive
molecule. Namely, chains-like structures for methanol($CH_{3}OH$)
(MetOH) and micelle-like structure for \emph{tert}butanol($(CH_{3})_{3}COH$)
(TBA). The first type of structure has been studied previously by
many authors and by various techniques\cite{guoSpectro,nartenMETH,sarkar}. 
It is often a matter
of debate as to know in which exact form these chain-like structures 
appear, such as rings, lassos and so on. In this work, we
would like to point out that the exact topology of the forms is not
so much relevant, since their enumeration is biased
by the theoretical description that is needed to spot
them out. Rather, the characterisation of the microstructure
by experimentally 
measurable quantities is the most relevant feature. 
Our observables are the two major static quantities related
to the structure mentioned above, namely an effective one body function,
as a \emph{local order parameter}, and the angle averaged pair correlation
function g(r), or equivalently the structure factor S(k) which is
the Fourier transform of g(r). The latter two functions are unambiguous
statistical quantities since they are directly measured 
by neutron or X-ray scattering experiments, or like in here calculated
by Molecular Dynamics simulations. 
Since the one body function is just the number
density of the system, a constant, we will here use a cluster counting
method in order to describe the degree of local heterogeneity in a
statistically meaningful way. For a given instantaneous configuration
$c$, the cluster of size $n$ is defined by counting all particles
that are paired within a given distance $l_{c}$. This is the
definition of the Stillinger cluster \cite{cluster}. 
In this way, we
have an instantaneous picture of the spatial distribution of the one
body density. By averaging over several such configurations the number
$s(n,c)$ of clusters of size $n$, we acquire a statistical picture
through the probability of having a cluster of size $n$: 
$p_{n}=\sum_{c}s(c,n)/\sum_{n}\sum_{c}s(c,n)$.
One can compute similarly a more precise quantity which is the probability
$p_{n}^{(XY)}$of having a clustering of some specific sites X and
Y on a molecule, and this is the quantity that we will consider here
as our local order parameter. By contrast to $g(r)$ or $S(k)$, $p_{n}$
and $p_{n}^{(XY)}$ have some degree of arbitrariness through the
choice of $l_{c}$. Despite this shortcoming, it turns out that both
type of quantities coherently point to the same microstructure of
neat alcohol. \\
Two models of MetOH are studied, the OPLS model \cite{OPLS} and the more recent
WS model \cite{smithMETH}, and for TBA the OPLS model \cite{OPLS}. Both
MetOH models have 3 sites, one for $O,H$, and one for the methyl group 
$M=CH_{3}$. Each site has a diameter and a partial
charge, and the interaction between 2 molecules is described as the
sum of the Lennard-Jones and Coulomb interactions between all pairs
of sites. Similarly, the TBA model is a six site model ($O,H,C,M_{1-3}$).
All the force fields used herein are classical atom-atom
force fields, and thus the hydrogen bonding (HB) is essentially electrostatic
in nature.
All simulations were performed 
in the constant NPT ensemble, with the Berendsen thermostat and barostat
to ensure a temperature of 300K and a pressure of 1 atm, with
a number of particles of N=2048. Rather than the radial distribution
function (RDF) between the center of masses of the molecules, we prefer
to use the site-site correlation functions $g_{a_{1}b_{2}}(r)$ between
the sites $a$ on molecule $1$ and site $b$ on molecule $2$ which
account better for the angular dependence of the correlations, as
well as the associated structure factors 
$S_{a_{1}b_{2}}(k)=1+\rho\int d\vec{r}\exp(-i\vec{k}.\vec{r})g_{a_{1}b_{2}}(r)$.
 Fig.1a shows some site-site RDFs for both MetOH models.
The methyl-methyl correlations has the oscillatory structure typical
of a Lennard-Jones (LJ) type liquid. In contrast, the correlations
involving the hydrogen bonding sites $O$ and $H$ display a striking
lack of structure at long range, after the first peak that is associated
with the HB tendency. We note that both models seem very close in
describing structural features. Fig.1b shows the site-site structure
factors corresponding to the functions in Fig.1a. Again, $S_{MM}(k)$
look typical of a LJ dense liquid, with a main peak around $k_{m}=2\pi/\sigma_{m}$,
where $\sigma_{m}$ is the effective diameter of the methanol molecule,
about $\sigma_{m}\approx 4.2\AA$ for both models. 
 \begin{figure}
 \includegraphics[width=9cm]{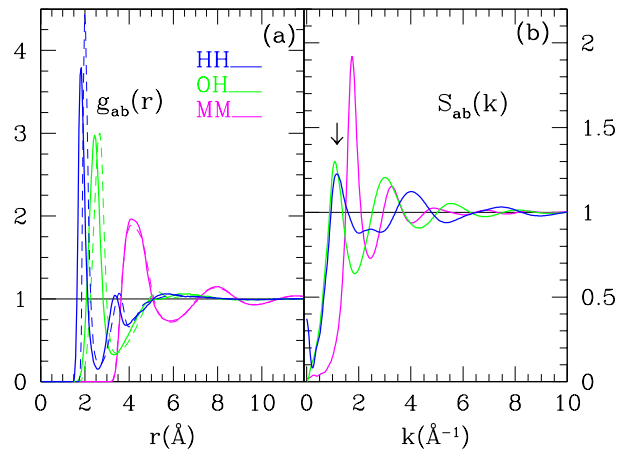}%
 \caption{\label{Fig.1} (color online) Density correlations for neat 
methanol (a) site-site g(r); full (OPLS), dashed (WS) (b) corresponding 
site-site structure factors (pre-peaks shown by arrow)}
 \end{figure}
The most important
feature in the structure factors involving the hydrogen bonding atoms,
is the presence of a pre-peak at k-vector $k_{p}\approx 1\AA^{-1}$
\emph{smaller} than $k_{m}$, which corresponds to a distance about
$r_{p}\approx6\AA$. 
This inner-peak is found in many reported experimental structure factors
\cite{nartenMETH,sarkar}.
It is an unmistakable trace of the microstructure of the neat liquid
due to hydrogen bonding.
Indeed, while hydrogen bonding corresponds
to a distance in the site-site RDF \emph{smaller} than $\sigma_{m}$-i.e.-
the position of the first peak in $g_{HH}(r)$ or $g_{OO}(r)$, 
and therefore 
should correspond to a \emph{larger} k-vector value, the
smaller $k_{p}$ value corresponds to a larger structure: the clusters 
formed by methanol molecules through the HB interactions. 
Can we recoup this interpretation
with the information from the order parameter, -i.e.- the cluster
probability? Fig.2a shows $p_{n}^{(XX)}$ , the probability of clustering
between similar sites $X$, as a function of the cluster size, for
the OPLS MetOH model. 
The inset shows the typical behaviour of $p_{n}^{(MM)}$
: there is a larger probability of finding smaller clusters, starting
from monomers, which decreases rapidly since larger clusters are less
probable. This situation is the same both for the methyl groups and
the hydrogen bonding sites. All of this is in accordance with the
intuitive idea of clustering in a liquid. 
A notable feature is the robustness of the general shape of $p_{n}^{(XX)}$ 
as a function of the distance $l_{c}$ that spans the well of the first
minimum of $g_{XX}(r) $.
Such plots have been shown in previous
studies by other authors. However, the main plot shows something that
has apparently not been noticed before: the small oxygen-oxygen cluster
size distribution $p_{n}^{(OO)}$ shows a small bump around $n\approx5$
\begin{figure}
 \includegraphics[width=9cm]{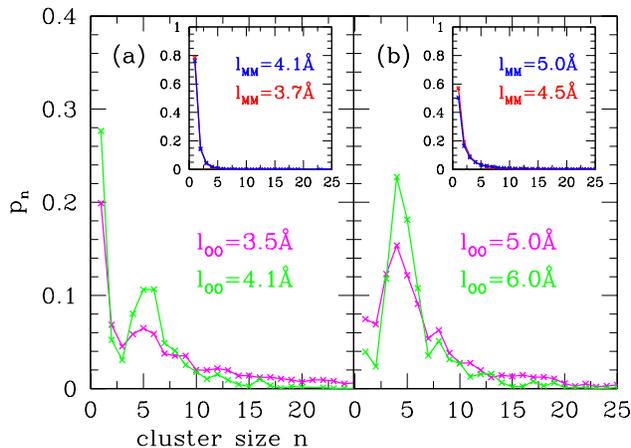}%
 \caption{\label{Fig.2} (color online) Cluster size probabilities 
for different sites (Oxygen and Methyl-group(insets)) and different 
cutoff values $l$ indicated in the plots;
(a) Methanol (OPLS) ; (b) {\it tert}butanol}
\end{figure}
that is not seen in $p_{n}^{(MM)}$. It indicates that about 5 oxygen
atoms cluster preferentially, through the HB mechanism, thus interlinking
the MetOH molecules, and in accord with the experimentally well known
fact that methanol molecules tend to form chains  
with rich topology (open chains, rings, lassos)\cite{guoSpectro,bates}. 
The absence of any bump in $p_{n}^{(MM)}$
is an indication of the topology of MetOH association: the connected oxygen and
hydrogen atoms form the backbone of the chain, 
while the methyl groups are randomly distributed around it. The
fact that monomer probability is higher than that for clusters corresponding
to the bump indicates that
the chain formation in liquid methanol is a weak feature. Returning
to the RDFs in Fig.1a, we now understand why methyl groups look packed
as in a monoatomic liquid, while the absence of correlation at the
large distance between hydrogen bonding sites is reminiscent of RDF
between monomers in polymeric fluids \cite{curro}, and thus indicates
chain formation between H-bonding sites.
\begin{figure}
 \includegraphics[width=9cm]{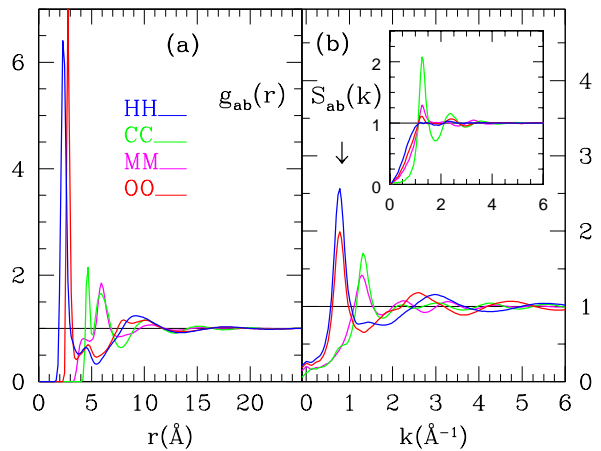}%
 \caption{\label{Fig.3} (color online) Density correlations for neat 
{\it tert}butanol (a) site-site g(r) (b) corresponding 
site-site structure factors. The inset shows the same data for 
uncharged sites}
\end{figure}

Turning to \emph{tert}butanol, Fig.3a shows the RDFs between some
particular sites. All RDFs have a strong first peak due to the hydrogen
bonding, but the oscillatory structure behind is different between
H-bonding and non-H-bonding sites: the latter have a period about the
average size of the molecule ($\approx 5.6\AA $) while the former have a
larger period of about $7.5 \AA$. This second period expresses the
modulation over the sphere-like suprastructure formed by the micelle-type
clustering of the TBA molecules. 
 Fig.3b shows the corresponding
S(k). Again, it is clear that while the non-hydrogen bonding sites
have an S(k) typical of an atomic liquid with $\sigma_{m}\approx5.7\AA$,
the other RDF show a pronounced pre-peak at $k_{p}\approx0.8\AA^{-1}$
that corresponds to a lengthscale of $r_{p}\approx7.5\AA$. In order
to confirm that this pre-peak is entirely due to the local organisation
coming from the hydrogen bonding tendencies between molecules, we
have simulated the TBA molecules without the partial charges, \emph{under
the same pressure and temperature}. The corresponding S(k) are shown
in the inset: the pre-peak have disappeared, and the whole atomic
distribution is as random as for the non-hydrogen bonding sites.
The clusters distribution is shown in Fig.2b. The methyl groups display
the random cluster distribution (inset) of ordinary liquids. It is
striking that the O-O cluster have a pronounced peak around size $n\approx 4$,
while the probability of finding monomers is now \emph{smaller} than
unity, indicating that liquid TBA is more strongly microstructured than methanol.
This fact is also transparent from the height of the pre-peaks in
Fig.3b, as compared to Fig.1b. 
 Inner peak and chain/ring type clustering in TBA have been reported from
various experiments \cite{tbaNarten,tbaSuhm}.
 A look at our simulations snapshots (Fig.5) indicates
that TBA molecules order by grouping all their hydrogen-bonding sites
together and letting the methyl tripods outside: thus forming small
micelles of about 4-6 molecules. \\
It is interesting to note that the electrostatic energy of methanol,
which accounts for the HB interaction, is $-29.60kJ/mol$ for the OPLS model
and $-35.4kJ/mol$ for the WS model, whereas it is $-24.73kJ/mol$ for the OPLS
model of TBA. The corresponding Van de Waals energies are respectively,
$-5.8kJ/mol$, $-6.1kJ/mol$ and $-21.kJ/mol$. 
Since the Coulomb energies are quite similar, the difference in patterns
is governed essentially by the topology of the molecule, in other
words by the symmetry of the interaction. The corresponding
experimental data are not available for the comparison, however
the experimental value for the enthalpy of liquefaction compares qualitatively
 with the calculated ones: for methanol $\Delta H=-37.3kJ/mol$ (expt),
$-35.44kJ/mol$ (OPLS) 
, while for TBA 
$\Delta H=-46.74kJ/mol$(expt) and $-45.74kJ/mol$(OPLS). 

\begin{figure}
 \includegraphics[width=8cm]{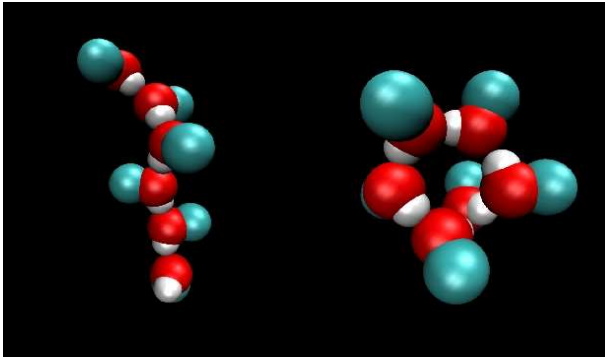}%
 \caption{\label{Fig.4} (color online) Typical local clusters
 of methanol molecules. Methyl in blue, oxygen in red and
hydrogen in white }
 \end{figure}
\begin{figure}
 \includegraphics[width=4.5cm]{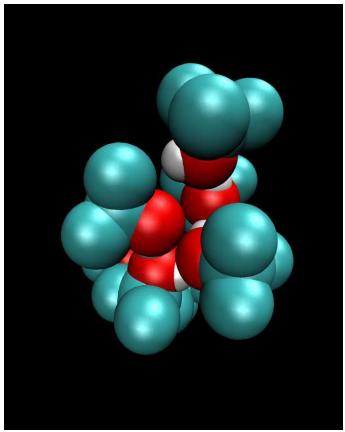}%
 \caption{\label{Fig.5} (color online) Typical local cluster
 of {\it tert}butanol molecules }
 \end{figure}
In view of the features reported for alcohols, 
 one may then ask what this type
of analysis would give for water-a stronger hydrogen bonding liquid.
The structure factor of water has been reported about three decades ago
\cite{waterNarten} and constantly improved since then.
It shows no sign of distinct pre-peak, except perhaps for a weak
shoulder at $k\approx2.0\AA^{-1}$. The cluster calculation indicates
no sign of bump or peaks in the small size region. These findings
indicate that water molecules do not form \emph{apparent}
clusters. One way to conciliate these inconsistencies with the present
approach is to conjecture that water molecules form branched-polymer
like clusters, due to the strong tetrahedrality of their hydrogen
bonds. Then, multiple configurations, due to the inherent
flexibility of such clusters, may not be 
detected by the cluster distribution and the pair correlations.
This conjecture remains to be tested. \\
The striking feature reported here is the dual appearance of homogeneity
and microheterogeneity in neat associated alcohols. They show
structural features strongly reminiscent of that taking place in microemulsion
\emph{after} the disorder to order phase transition, while present
features are \emph{within} the disordered phase. This association
is visible in the pair correlations, particularly in the structure
factors, but also in the one-body distributions modified to account
for local structures. As said in the beginning, while we understand
that these features are related to the strong directionality of the
hydrogen bonding tendency in these liquids, we cannot help being surprised
that this interaction creates equilibrium heterogeneity in a globally
homogeneous \emph{disordered} liquid.
The view point introduced herein
is that associated liquids can be dually viewed as constituted of
molecules interacting through strong directional forces and at the
same time as mixtures of microclustered molecular domains. The implications
in the physical chemistry of such liquids and their mixture with water
remains an open problem.


\begin{thebibliography}{10}
\bibitem{soperNAT}S. Dixit, J. Crain, W.C.K. Poon, J. L. Finney and
A.K. Soper, Nature \textbf{416}, 6883 (2002)


\bibitem{soperPERCO}L. Dougan, S. P. Bates, R. Hargreaves, J. Fox,
J. Crain, J. L. Finney, V. Reat and A. K. Soper, J. Chem. Phys. \textbf{121},
6456 (2004)

\bibitem{bates}S. K. Allison, J. P. Fox, R. Hargreaves and S. P.
Bates, Phys. Rev. \textbf{E71}, 24201 (2005)

\bibitem{guoSpectro}J.-H. Guo, Y. Luo, A. Augustsson, S. Kashtanov,
J.-E. Rubensson, D. K. Shuh, H. \AA gren and J. Nordgren, Phys. Rev.
Lett. \textbf{91}, 157401-1(2003)

\bibitem{aupKB}A. Perera, F. Sokoli\'c, L. Alm\'asy and Y. Koga,
J. Chem. Phys. \textbf{124}, 124515 (2006)

\bibitem{smithMETH} S. Weerasinghe and P. E. Smith, J. Phys. Chem.
\textbf{B109}, 15080 (2005)

\bibitem{nico}M. E. Lee and N. F. A. van der Vegt, J. Chem. Phys.
\textbf{122}, 114509 (2005)

\bibitem{frank}H. S. Frank and A. S. Quist, J. Chem. Phys. \textbf{34},
604 (1960)

\bibitem{stanleyWaterMyst} H. E. Stanley et al.
J. Phys. Cond. Mat. \textbf{12}, A403 (2000)

\bibitem{debenedetti}J. R. Errington, P. G. Debenedetti and S. Torquato,
Phys. Rev. Lett. \textbf{89}, 215503-1 (2002)

\bibitem{stanleyPERCO}H. E. Stanley, J. Teixeira, A. Geiger and R.
L. Blumberg, Physica \textbf{A106}, 260 (1981)

\bibitem{nartenMETH} A. H. Narten and A. Habenschuss, J. Chem. Phys. 
\textbf{80}, 3387 (1984)

\bibitem{sarkar}S. Sarkar and R. N. Joarder, J. Chem. Phys. \textbf{99},
2032 (1993)

\bibitem{cluster}L. A. Pugnaloni and F. Vericat, J. Chem. Phys. \textbf{116},
1097 (2002)

\bibitem{OPLS} W. L. Jorgensen, J. Phys. Chem. 90, 1276 (1986)

\bibitem{curro} J. G. Curro and K. S. Schweizer, J. Chem. Phys. 87, 1842 (1987)
\bibitem{tbaNarten} A. H. Narten and S. I. Sandler, J. Chem. Phys. \textbf{71},
 2069 (1979)
\bibitem{tbaSuhm} D. Zimmerman, TH. H\"{a}ber, H. Schaal and M. A. Suhm,
 Mol. Phys. \textbf{99}, 413 (2001)
\bibitem{waterNarten} A. H. Narten and H. A. Levy, J. Chem .Phys. \textbf{55},
2263 (1971)

\end{thebibliography}
\end{document}